\documentclass[%
 reprint,
 amsmath,amssymb,
 aps,
]{revtex4-2}

\usepackage{graphicx}
\usepackage{dcolumn}
\usepackage{bm}
\usepackage{hyperref}
\usepackage[mathlines]{lineno}
\usepackage{xcolor}

\begin{document}

\preprint{APS/123-QED}

\title{Negative Temperature Pressure in Black Holes}

\author{Richard A. Norte}
 \affiliation{Delft University of Technology, Mekelweg 2, 2628CJ Delft, The Netherlands}

\date{\today}

\begin{abstract}
The concept of negative temperature (T $<$ 0) is unique to quantum physics and describes systems that are hotter than any positive temperature system. For decades negative temperatures have been shown in a number of spin systems, but experiments only recently demonstrated atomic ensembles with negative temperatures in their motional degrees of freedom. An observed behavior of such negative temperature ensembles is that despite highly attractive forces between an arbitrary number of particles, there is a self-stabilization against collapse. Negative temperatures are only possible in quantum systems because there exists upper bounds on the energy of particles -- a property not found in classical physics. Here we consider whether event horizons set up similar upper limits within black holes, giving rise to negative temperature systems just within event horizons. Combining black hole thermodynamics with experimentally observed negative temperature effects could imply a quantum-based outward pressure in black holes. \\
\end{abstract}

\maketitle  

Stellar objects are a ubiquitous example of quantum effects balancing robustly against gravity. In the Sun, quantum effects are responsible for the nuclear fusion reactions that stave off gravity's inward push. In ultra-dense stars like white dwarfs and neutron stars, quantum effects, like electron and neutron degeneracy pressure, provide the outward force countering gravity's pull (Fig. 1). This balance is defined by critical mass limits such as the Chandrasekhar limit for white dwarfs~\cite{chandrasekhar1931maximum} and the Tolman-Oppenheimer-Volkoff limit for neutron stars~\cite{oppenheimer1939massive}; similarly quark stars have also been suggested~\cite{quarkstar1965}. An unusual exception to this balancing act are black holes, where gravity is thought to completely dominate~\cite{penrose1965gravitational,hawking1973large} resulting in a singularity - a point-mass of infinite density. Experimental approaches have recently looked towards atomic ensemble experiments as analogue gravitation systems to illuminate the underlying physics of black holes. To date, atomic ensembles have been used to detect Hawking Radiation from ultra-cold condensates~\cite{munoz2019observation,kolobov2021observation}, and to simulate Unruh radiation~\cite{hu2019quantum} and event horizons~\cite{franz2018mimicking, oppenheim2003}. While our current understanding of physics may not be sufficient to describe the interiors of black holes, it is crucial to consider that these regions may involve quantum gravitational effects, potentially allowing for unconventional behaviors like negative temperatures found only in quantum systems. Black holes are generally considered cold objects near absolute zero ($T = +0$), with ultra-low entropy that uniquely scales with the surface area of the event holes. In this paper, we explore the possibility that black holes could harbor conditions for `local' negative temperatures ($T = -0$) just within the event horizon, generating an outward pressure within a black hole — a mechanism inspired by recent landmark experiments.

\begin{figure}[t!]
	\begin{center}
		\includegraphics[width=\columnwidth]{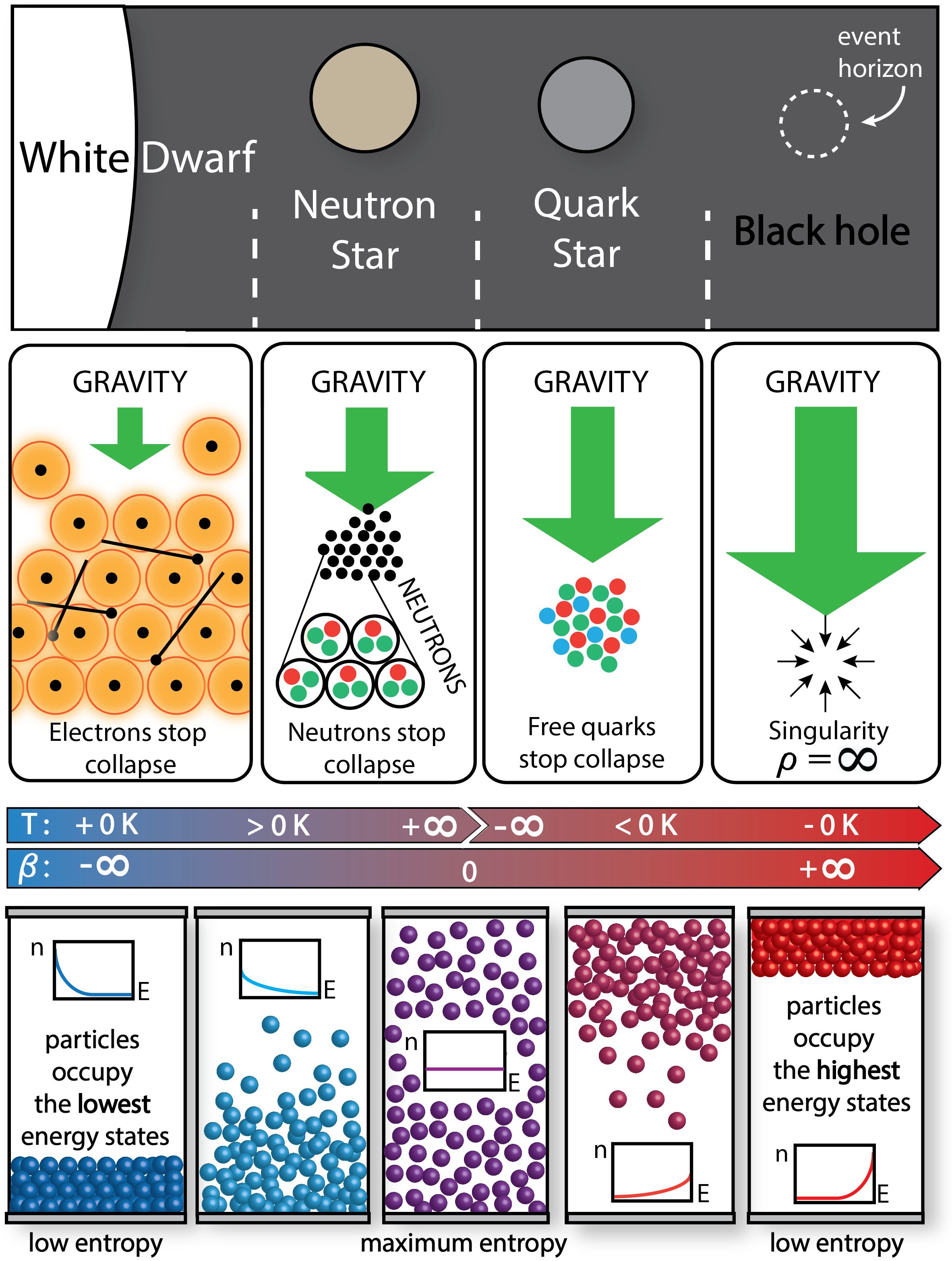}
		\caption{\textbf{Top:} Illustration of gravity counteracted by quantum degeneracy pressure within a white dwarf, neutron star, and quark star. Conventionally gravitational collapse form a singularity within a black hole's event horizon. \textbf{Bottom:} Schematic based on Ref.~\cite{braun2014thesis} shows the transition from absolute zero to negative temperatures in terms of T, $\beta$, and the particle number, n, which occupy states of energy E in the system. We consider whether the formation of event horizons set up analogous upper bounds within black holes.} 
		\label{fig:S7}
	\end{center} 
\end{figure}

This manuscript is structured as follows: We commence by explaining negative temperatures, focusing on a 2013 experiment that observed these in motional degrees of freedom in atomic ensembles and identified a stabilizing outward pressure. Then we hypothesize that similar conditions might manifest within black hole event horizons. To substantiate this, we relate ultra-cold atomic ensembles to black holes, emphasizing that insights gained in one context could be instructive for the other. Using a theoretical framework that approximates black holes via Ising models, we describe mechanisms through which negative temperatures could emerge within an event horizon. We conclude by discussing the implications for black hole thermodynamics. 

\subsection{Negative Temperatures}
Temperature is conventionally seen as the average kinetic energy of the system's particles. Negative temperatures are a uniquely quantum occurrence which allows for low entropy systems which are hotter than an infinitely hot system. This oddity can be understood by considering the more strict definition of thermodynamic temperature as the change of entropy with energy change in a system $\partial S/\partial E = 1/T$. In such descriptions, `coldness', the reciprocal of temperature ($\beta = 1/k_B T$), is the more fundamental quantity. Systems with a positive temperature will continuously increase in entropy as one adds energy to the system, while systems with a negative temperature will decrease in entropy as one adds energy to the system (as shown at the bottom of Figure 1). 

Negative temperatures first emerged in quantum spin systems under magnetic fields. In these systems, all spins align with the magnetic field at close to absolute zero ($T = +0K$, $\beta = -\infty$), resulting in minimal entropy. Upon heating, the spins gradually oppose the magnetic field, and entropy increases. As the system approaches infinite temperature ($T \rightarrow \pm \infty$, $\beta \rightarrow 0$), the spins become evenly distributed between energy states, maximizing entropy. Beyond this point, a regime of negative temperatures ($T < 0K$, $\beta > 0$) occurs when more spins populate the higher energy state, thus reducing entropy. The system reaches its lowest entropy again when all spins oppose the magnetic field at near-absolute negative temperature ($T \rightarrow -0K$, $\beta \rightarrow +\infty$). Negative temperatures are specific to quantum systems because classical systems lack an upper energy bound and can continually absorb energy. 

While negative temperatures in spin systems have been studied for decades, a groundbreaking 2013 experiment first demonstrated negative temperatures in particles' motional degrees of freedom, rather than the traditional spin degrees. Remarkably, this system resisted inward collapse regardless of particle count, even with strong attractive forces between particles. The experiment revealed that negative temperatures inherently yield negative pressures, which counteract an expected collapse. This phenomenon is explained through basic thermodynamics by Braun et al~\cite{braun2013}, as summarized in the following section.  
   
\subsection{2013 Negative Temperature Experiments}

Braun et al.~\cite{braun2013} studied an ensemble of bosons with negative temperature in their motional degrees of freedom. Using a tailored Bose-Hubbard Hamiltonian, they managed to stabilize an attractively interacting ensemble against collapse for any particle number. To explain this unexpected stability, they developed a generalized stability condition rooted in thermodynamics.

In thermal equilibrium, a gas aims to maximize its entropy \( S \) under given constraints, like fixed total energy \( E \) and maximum volume \( V_{\text{max}} \). Usually, this results in a positive pressure \( P \), filling the entire available volume. The stability condition they propose simplifies to:
\begin{equation}
\frac{\partial S}{\partial V} \bigg|_E \geq 0
\end{equation}
A negative value for this derivative would indicate that the gas could increase its entropy by contracting, making it unstable against collapse. This condition can be derived from the energy differential:
\begin{equation}
dE = T~dS - P~dV 
\end{equation}
leading to:
\begin{equation}
\frac{\partial S}{\partial V} \bigg|_E = \frac{P}{T} 
\end{equation}
In equilibrium, \( P \) and \( T \) must have the same sign, meaning that negative temperatures imply negative pressures and vice versa. Braun et al.'s experiment demonstrates that a negative temperature can actually stabilize an attractively interacting ensemble, preventing it from collapsing.

\subsection{Negative Temperatures in Black Holes}

We investigate the role of an event horizon as a potential upper bound for the Hamiltonian inside the Schwarzschild radius, and thereby assess its implications for negative temperatures. Although the concept of negative temperature within black holes has precedent in the literature~\cite{oppenheim2003, tiandho2017implication, bo2010negative, park2007thermodynamics, cvetivc2018killing}, we utilize Jonathan Oppenheim's lattice model~\cite{oppenheim2003} as our theoretical framework. This model treats black holes as gravitational analogues of Ising-like lattice systems with long-range interactions, comparable to atomic ensembles in experiments. Importantly, this approach shows that systems whose entropy scales with area are not unique to black holes but can also occur in lattice systems with long-range interactions. Such non-extensive entropy observed in lattice models with long-range interactions can offer insights into the entropy characteristics of black holes, suggesting similar non-extensive properties due to their inherent long-range gravitational interactions. Exterior to black holes, Oppenheim's model reproduces Hawking's well-known predictions~\cite{hawking1975particle} such as black hole temperature and entropy for a distance observer. We first demonstrate the possibility of negative temperatures within an event horizon using the lattice model. This would imply that interiors of black holes are hotter than is possible for any classical systems, yet can harbor low-entropy expected of black holes. Subsequently, we discuss how pressure and volume are treated in the first law of black hole thermodynamics. Our aim is to align these with the thermodynamic stability conditions in Eq.~1-3 to explore the possibility of `negative temperature pressure' within black holes.

\subsection{Lattice Models with Long-Range Gravitational Coupling}

We utilize Jonathan Oppenheim's work \textit{Thermodynamics with long-range interactions: From Ising models to black holes}~\cite{oppenheim2003}. Oppenheim models black holes as lattices of interacting spins, paralleling the Ising model but incorporating gravitational interactions. In this formulation, the lattice system is characterized by the total energy \( m \), which depends on the magnetic fields \( h_j \) and spin-spin interactions \( J_{jk} \). Consider a lattice of N spins with total energy

\begin{equation}
m = \sum_{j} h_j \sigma_j - \sum_{<j,k>} J_{jk}\sigma_j \sigma_k
\end{equation} where $h_j$ are the magnetic field values and $\sigma$ is spin at the lattice site which is $\pm 1$. Now assume spin-spin coupling is strong enough for $J_{jk}$ and $h_j$ to be almost constant. Assuming spin-spin coupling and magnetic fields to be nearly uniform over large spatial scales, Oppenheim simplifies the system's dynamics, arriving at an equation for the total energy \( m \) to 

\begin{equation}
m = he - J e^2/2
\end{equation}where the local energy is $E = he$. There are two possible values for the local energy. Solving for e, 

\begin{equation}
e_{\pm}(m) = \frac{h}{J} [1 \pm k(m)]
\end{equation}
where
\begin{equation}
k(m) = \sqrt{1-\frac{2Jm}{h^2}}
\end{equation}

Within this framework, he introduces both a global and a local temperature. The global temperature serves as an `average' thermal metric for the entire system, while the local temperature provides a measure that can vary depending on the local energy states. This global temperature is also written in terms of total energy and interacting terms. Inspired by general relativity, Oppenheim also describes a local temperature measured in a local frame. Both temperatures are physically measurable. 

The conventional derivation of a canonical ensemble considers a large reservoir $R$ in contact with a smaller system $S$. One fixes the total energy of the combined system (hence, one is operating in the microcanonical ensemble), but also allow energy flow between $R$ and $S$. The distribution depends on a quantity which is defined as the temperature and this defines the canonical ensemble. It can be thought of as a system in a probability distribution of different canonical ensembles. Alternatively it is possible to formulate the micro-canonical distribution, but it proves helpful to define a new ensemble which Oppenheim calls the microlocal ensemble. Rather than fixing the total energy $m$, they fix the total local energy $E = E_R + E_S$. Global temperature is mathematically
defined as
\begin{equation}
\beta_0 = \frac{\partial E}{\partial m}\beta_E
\end{equation}
where $\beta_E$ is the local temperature. To get the relationship between global and local temperature one can combine $e_{\pm}$ and $\beta_0$ to show that 
\begin{equation}
\beta(e_{\pm}) = \beta_0 \left(1-\frac{J e_{\pm}}{h}\right) = \mp \beta_0 \sqrt{1-\frac{2mJ}{h^2}} 
\label{mJEQ}
\end{equation}

What's particularly compelling about Oppenheim's approach is how closely it parallels general relativity. By simply mapping the magnetic field parameter \( h^2 \) to \( r \) and the spin-spin coupling \( J \) to the gravitational constant \( G \), Oppenheim bridges the Ising model and its gravitational analogue. His model produces results that are surprisingly similar to the Tolman relations, which in general relativity describe how temperature varies in a curved space-time setting. In this model, the local temperature \( \beta(e_{\pm}) \) is modulated by the local energy \( e_{\pm} \) and the remapped parameters \( G \) and \( r \). The relation 
\begin{equation}
\beta_{\pm} = \mp \beta_0 \sqrt{1-\frac{2GM}{r}}
\end{equation}
reveals how local temperatures can vary throughout the system, influenced by the total energy \( m \) and the model's gravitational analogues. This variation even mimics the red-shifting of temperature observed in the exterior of a Schwarzschild black hole. Following Oppenheim's framework, we propose that the internal regions of a black hole might not exhibit a uniform temperature. Instead, a black hole could be characterized by a distribution of local temperatures.

Even more intriguingly, Oppenheim's model suggests the existence of negative local temperatures within the event horizon of a black hole analogue where $\beta(e_{\pm}) =\mp \beta_0 \sqrt{1-2Gm/r}$ . This result is grounded in general relativity, where the tilting of the light cone inside a black hole effectively turns what appears as positive energy from an external viewpoint into negative energy internally. As a result, spins within this black hole analog are more likely to inhabit states corresponding to these negative local temperatures.

\subsection{Negative Temperature Pressure in Black Holes}

In the case of a Schwarzschild black hole (non-rotating, uncharged), the first law of black hole thermodynamics can be stated as:
\begin{equation}
    dE = T dS + \Phi dQ - \Omega dJ
\end{equation}

In the Schwarzschild case, the last two terms are zero (\(dQ=0\) and \(dJ=0\)) since the black hole is uncharged and non-rotating, simplifying the equation to:

\begin{equation}
    dE = T dS
\end{equation}

Now, if we consider that the black hole's interior has a negative temperature (\(T_i < 0\)), then we can make thermodynamic arguments related to stability. The entropy \(S\) of the black hole is proportional to the area \(A\) of the event horizon,
\begin{equation}
    S = \frac{kA}{4}
\end{equation}
where \(k\) is a constant of proportionality. We can then relate the change in entropy to a change in the radius \(r\) of the black hole:

\begin{equation}
    dS = k dA = k (8\pi r dr) = 8\pi k r dr
\end{equation}

Then, we can relate this to a change in the mass-energy \(E=Mc^2\) of the black hole. Assuming that \(dE\) is also negative, we have:

\begin{equation}
    dE = -|T_i| dS
\end{equation}

To explore the implications for stability, let's introduce a concept analogous to `pressure' for the black hole, defined as \(P_{\text{BH}}\). The stability condition would then be:

\begin{equation}
    \left( \frac{\partial S}{\partial r} \right)_E \geq 0 
\end{equation}

This condition, much like in the case of the bosonic ensemble, can be derived from the first law of black hole thermodynamics:

\begin{equation}
    \left( \frac{\partial S}{\partial r} \right)_E =  \frac{P_{\text{BH}}}{T_i} 
\end{equation}

In thermal equilibrium, \(P_{\text{BH}}\) and \(T_i\) must have the same sign, meaning that a negative temperature would imply a negative ``pressure'' within the black hole. A negative pressure would generate an outward force against the gravitational pull, potentially stabilizing the black hole against further collapse. Large black holes where Hawking radiation is negligible compared to the mass-energy have been considered in thermal equilibrium~\cite{ling2022anti,page2005hawking}.   

Thus, if a black hole's interior could have negative temperatures, similar logic suggests that this could create a negative pressure inside the black hole, preventing it from collapsing indefinitely. It is important to note that the meaning of thermal equilibrium, pressure, volume in black hole thermodynamics are lines of active research in physics which we briefly discuss in the next section. 

\subsection{Pressure and Volume in Black Hole Thermodynamics}

The formulation of negative temperatures by Braun et al., given in Eq.~1 and Eq.~2, is based on straightforward thermodynamic reasoning, using the energy differential \(dE = T~dS - P~dV\). Specifically, incorporating a similar \(P~dV\) term in the realm of conventional black hole thermodynamics remains an open question.

According to Ref.~\cite{Dolan12}, the first law of black hole thermodynamics can be expressed in terms of the mass \(M\) of the black hole, identified as its internal energy \(U(S)\). The equation becomes \(dU = T~dS\). For electrically charged or rotating black holes, it generalizes to
\begin{equation}
    dU = T~dS + \Omega~dJ + \Phi~dQ,
\end{equation}
where \(J\) is the angular momentum, \(\Omega\) is the angular velocity, and \(Q\) is the electric charge. Notably, the equation lacks the \(P~dV\) term commonly found in standard thermodynamics.

Recent theories propose interpreting the cosmological constant, \(\Lambda\), as a thermodynamic variable representing pressure. This has been extensively explored~\cite{Dolan12}, although a clear physical interpretation of pressure or volume has not yet been established. Ref.~\cite{kastor2009enthalpy} argues for the inclusion of \(\Lambda\) in thermodynamic variables to maintain consistency with the Smarr relation. According to this work, the mass of a black hole in anti-de Sitter (AdS) space should be interpreted as the enthalpy \(H\) rather than the internal energy \(U\),
\begin{equation}
M = H(S,P) = U(S,V) + PV.
\end{equation}

This allows us to write the first law of black hole thermodynamics as
\begin{equation}
    dU = T~dS + \Omega~dJ + \Phi~dQ - P~dV,
\end{equation}
where \(P\) and \(V\) have particular interpretations in the context of AdS space-time. For Schwarzschild black holes, the equation simplifies to
\begin{equation}\label{bhthermo}
    dU = T~dS - P~dV.
\end{equation}

This formulation aligns well with Braun et al.'s stability formulation, facilitating the same thermodynamic reasoning for negative temperatures and pressures. However, the physical interpretation of pressure and volume in black hole thermodynamics should be explored further. \newline

\subsection{Conclusions}
Recent experiments~\cite{braun2013} have provided compelling evidence for particles with negative temperatures in their motion degrees of freedom. We examined the potential for such negative temperatures to exist within black holes and proposed that this could generate an outward pressure countering gravitational collapse. Using Oppenheim's formulation~\cite{oppenheim2003}, we consider black holes as gravitational lattice ensemble with strong long-range interactions to arrive at negative temperatures, and subsequently incorporate a \(PdV\) term in black hole thermodynamics to show how outward negative pressures can arise within; analogous to experimental observations with atomic ensembles. The exterior black hole thermodynamics remain consistent with conventional theory~\cite{oppenheim2003}. Several questions arise from this hypothesis, including the interpretation of pressure and volume in black hole thermodynamics and the viability of a stable structure within the event horizon. If stable, this would present an interior black hole structure that paradoxically uses intense gravity to maintain its form, all while aligning with Hawking's thermodynamics externally. Future research could investigate the potential for observing this negative temperature behavior through gravitational wave signals from inspiral black hole collisions~\cite{cardoso2019testing, mazur2015surface, pani2010gravitational} or analogue gravitational systems in atomic ensembles~\cite{munoz2019observation,kolobov2021observation}. A recent study\cite{annala2023strongly} demonstrated that neutron-star cores likely undergo a phase transition to deconfined quark matter, with the equation of state (EoS) in the cores of massive stars consistent with quark matter. This finding was achieved by combining astrophysical observations with theoretical analysis, using Bayesian inference to establish conformal symmetry at high densities indicative of deconfined matter. Similar approaches could be employed to probe the interiors of black holes for signatures of negative temperatures and potentially distinct EoS. Understanding this phenomenon could offer novel insights into the interplay between quantum pressure and gravity, as recently indicated by studies showing that non-local quantum effects can generate a pressure for black holes~\cite{calmet2021quantum}. Negative temperature states are usually enabled by electromagnetic forces; this work raises the possibility that gravity could serve an analogous role in black holes. Conventional views suggest black holes approach absolute zero from the positive temperature side \(T \rightarrow +0\); our model suggests they might also approach from the negative side \(T \rightarrow -0\), providing a unique mechanism for negative temperature pressures. This could point to a novel quantum-gravitational interaction that potentially make black holes similar to other stellar objects - a balance between quantum effects and gravity. 
\\
\begin{acknowledgments}
I would like to acknowledge the inspiring work of Braun et al., Jonathan Oppenheim, and Brian Dolan whose derivations and clear explanations were directly utilized to convey the paper's main concepts. I would also like to thank Faizal Mir, Yanbei Chen, and James Q. Quach for interesting discussion and feedback.
\end{acknowledgments}

\appendix

\nocite{*}

\bibliography{References}

\end{document}